\documentclass[twocolumn,showpacs,preprintnumbers,amsmath,amssymb]{revtex4-1}

\usepackage{cancel}
\usepackage{graphicx}
\usepackage{dcolumn}
\usepackage{bm}
\usepackage{color}
\def\br#1{\left( #1 \right)}

\def\brac#1{\left[ #1 \right]}

\def\Tr{{\rm Tr}}
\def\non{\nonumber}

\begin{document}

\preprint{PKNU-NuHaTh-2020-02}

\title{
The use of the canonical approach in effective models of QCD
}

\author{Masayuki~Wakayama$^{1,2,3}$}
\email{wakayama@rcnp.osaka-u.ac.jp}
\author{Seung-il~Nam$^{1,2,4}$}
\author{Atsushi~Hosaka$^{3,5}$}
\affiliation{$^1$Department of Physics, Pukyong National University (PKNU), Busan 48513, Republic of Korea}
\affiliation{$^2$Center for Extreme Nuclear Matters (CENuM), Korea University, Seoul 02841, Republic of Korea}
\affiliation{$^3$Research Center for Nuclear Physics (RCNP), Osaka University, Ibaraki, Osaka 567-0047, Japan}
\affiliation{$^4$Asia Pacific Center for Theoretical Physics (APCTP), Pohang 790-784, Republic of Korea}
\affiliation{$^5$Advanced Science Research Center, Japan Atomic Energy Agency (JAEA), Tokai 319-1195, Japan}

\date{\today}

\begin{abstract}
We clarify regions where the canonical approach works well at the finite temperature and density 
in the Nambu-Jona-Lasinio (NJL) and Polyakov-NJL (PNJL) models. 
The canonical approach is a useful method for avoiding the sign problem in lattice QCD simulations at finite density, 
but it involves some parameters. 
We find that number densities computed from the canonical approach are consistent with exact values in most of the confinement phase 
within the parameters, which are applicable in lattice QCD. 
\end{abstract}

\pacs{
}

\maketitle

\section{\label{intro}Introduction}

Understandings for Quantum Chromodynamics (QCD) at finite temperature and density 
have been highly demanded to fundamental inputs in various interesting questions 
such as 
the generation of matter in the early universe, the galaxy formations and mysterious stellar objects 
such as neutron stars and black holes. 
The high energy accelerators at such as J-PARC (KEK/JAEA), FAIR (GSI) and NICA (JINR) 
will be expected to operate in the near future as experimental approaches to the questions. 
In the theoretical side, it is well known that lattice QCD is an almost unique method for the first principle simulations of QCD. 

As already well known, however, lattice QCD simulations suffer from the sign problem at finite density. 
The canonical approach~\cite{Hasenfratz:1991ax}, which is one of the methods proposed to avoid the sign problem, 
has been developed rapidly 
with multiple-precision arithmetic~\cite{Morita:2012kt,Fukuda:2015mva,Nakamura:2015jra,deForcrand:2006ec,
Ejiri:2008xt,Li:2010qf,Li:2011ee,Danzer:2012vw,Gattringer:2014hra,
Boyda:2017lps,Goy:2016egl,Bornyakov:2016wld,Boyda:2017dyo,Wakayama:2018wkc,Wakayama:2019hgz}. 
The canonical approach can be applied to study the physical observables such as particle number distributions in heavy-ion collisions 
and reveal the phase structure at $\mu$ similar to the effective quark mass $\sim300$~[MeV] for the light-flavor $SU(2)$ sector. 
However, there is a question of the validity of the method when the lattice data that can be used for the analyses is limited. 

In this paper, we would like to address this question by using QCD effective models 
such as the Nambu--Jona-Lasinio (NJL) and Polyakov-loop augmented NJL (PNJL) ones. 
The advantage of the models is that it is possible to perform (semi) analytically the canonical approach. 

The NJL model has been successful in describing various properties of nonperturbative QCD~\cite{Nambu:1961tp,
Nambu:1961fr,Kunihiro:1991qu,Hatsuda:1994pi}. 
In our previous paper~\cite{Wakayama:2019hgz}, 
the model was applied to the Lee-Yang zero problem of the QCD phase structure. 
The PNJL model incorporates not only spontaneous symmetry breaking of chiral symmetry 
but also the spontaneous breaking of $Z(N_c)$ symmetry. The latter is governed by 
the expectation value of the Polyakov loop $\langle\Phi\rangle$ 
as an order parameter for confinement and deconfinement phases~\cite{Fukushima:2003fw,Rossner:2007ik}. 
In this way, the PNJL model incorporates partly the gluon dynamics. 

Our strategy is as follows. 
At real finite chemical potentials, 
we cannot perform lattice QCD simulations due to the sign problem caused by complex values of the grand canonical partition function. 
In the canonical approach, lattice QCD is calculated at pure imaginary chemical potentials 
where the grand canonical partition function is real, that avoids the sign problem. 
In accordance with the lattice data analysis, 
first, we compute the quark number density at pure imaginary chemical potentials in the effective models. 
The resulting quark number density as a function of the chemical potential is parametrized 
by a Fourier series of a finite number of terms $N_{\sin}$.  
The validity of the canonical approach is determined by the accuracy of the parametrization, 
the investigation of which is the main subject of the present paper.  
Furthermore, we introduce the maximum value of fluctuations of the net quark number $N_{\rm max}$ 
that is needed in lattice simulations due to finite amounts of resources. 
A comparison of the results of finite $N_{\rm max}$ with the exact ones 
also provides a measure of the validity of the canonical approach in the actual lattice simulations.

From the numerical results, 
we find that the canonical approach works qualitatively well even near the phase-transition line for relatively small values of 
$N_\mathrm{max}$ and $N_\mathrm{sin}$, 
$N_\mathrm{max}/V \gtrsim 0.56$~[fm$^{-3}$] and $N_\mathrm{sin}\approx4$, 
where $V$ is a volume in the system. 
Especially, $N_{\sin}=1$ or 2 is enough to reconstruct the exact number density within the 10\% difference from the canonical approach 
for the temperature below $T^\mathrm{CEP}$ and $\mu_B$ below about 900~[MeV].

The present paper is organized as follows: 
In Section II, we briefly explain the canonical approach in the PNJL model. 
The numerical results are given in Section III with detailed discussions. 
Section IV is devoted to summary and future perspectives.

\section{\label{cano}The canonical approach in the PNJL model}
\subsection{The canonical approach}

In this subsection, we review the canonical approach. 
First, there is a relation between the grand canonical partition function $Z_{GC}$ and the canonical partition functions $Z_{C}$ 
as a fugacity expansion, 
\begin{eqnarray}
Z _{\rm GC} (\mu,T,V) 
&=& \sum_{n=-\infty}^{\infty} Z_C(n,T,V) \xi^n \ , \label{fuga_exp}
\end{eqnarray}
where $\mu$, $T$, $V$ and $\xi(\equiv e^{\mu/T})$ are 
the quark chemical potential, temperature, volume of the system and the quark fugacity, respectively. 
The Fourier transforms of Eq.~\eqref{fuga_exp} can be written as 
\begin{eqnarray}
\!\!\!\!\!\!\!\!
Z_C(n,T,V)  &=& \int_{-\pi}^{\pi} \frac{d\theta}{2\pi} \, e^{-in\theta} Z_{\rm GC} (\mu=i \mu_{I},T,V) \ , \ \ \ 
\label{ZC}
\end{eqnarray}
where $\mu_I$ is real and $\theta = \mu_{I}/T$. 
Because the Fourier transforms have cancellations of significant digits 
that come from the high frequency of $e^{-in\theta}$ at large $n$, 
multiple-precision arithmetic is needed in numerical calculations. 

Furthermore, the integration method is used to extract $Z_C$ for large $n$ 
in lattice QCD calculations~\cite{Boyda:2017lps,Goy:2016egl,Bornyakov:2016wld,Boyda:2017dyo,Wakayama:2018wkc}. 
In the integration method, 
$Z_{GC}(i\mu_{I})$ in Eq.~\eqref{ZC} is derived from the number density at the pure imaginary chemical potential, 
\begin{eqnarray}
 \frac{n_{q}}{T^3}(i\mu_I) &=& \frac{1}{VT^2} \frac{\partial}{\partial (i\mu_I)} \ln Z_{GC}(i\mu_I) \ . 
 \label{integration}
\end{eqnarray}
Because $Z_{GC}(i\mu_{I})$ is real, 
we can define as $n_q(i\mu_I)=in_{qI}$ with the real valued $n_{qI}$. 
The imaginary number density $n_{qI}$ is well known to be approximated by a Fourier series, 
\begin{eqnarray}
 \frac{n_{qI}}{T^3}(\theta) &=& \sum_{k=1}^{N_{{\sin}}} f_{k} \sin(k\theta) \ , 
 \label{mulsin}
\end{eqnarray}
with 
a finite number of terms of $N_{\sin}$~\cite{DElia:2009pdy,Takaishi:2010kc}. 
After getting a set of coefficients $f_{k}$, 
we can evaluate $Z_{GC}(i\mu_I)$ in good approximation from 
\begin{eqnarray}
\!\!\!\!\!\!\!\!\!\!\!\!
Z_{\rm GC} (i \mu_{I},T,V) 
&=& C \exp\left[-V \int^{\theta}_0 d\theta^{\prime} \, n_{qI}(\theta^{\prime})\right] \non \\
&=& C \exp\left[VT^3 \sum_{k=1}^{N_{\sin}} \frac{f_k}{k} \cos \br{k\theta}\right]  \ , \ \ \ \ \ 
\label{GCint}
\end{eqnarray}
where $C$ is an integration constant. 

\subsection{The PNJL model}

The effective potential $\omega$ of the PNJL model 
is given as 
\begin{eqnarray}
\omega &=&\frac{1}{2G} \br{M-m_q}^2 
                                              - 2N_c N_f \int \frac{d^3p}{\br{2\pi}^3}   E_{p}  \non \\
&& -2 N_f T \int \frac{d^3p}{\br{2\pi}^3} \Big\{
                                                \Tr_c  \ln\brac{1+ L e^{-\frac{E_p-\mu}{T}}} \non \\
                          && \ \ \ \ \ \ \ \ \ \ \ \ \ \ \ \ \ \ \ \ 
                               +  \Tr_c  \ln\brac{1+ L^{\dag} e^{-\frac{E_p+\mu}{T}}}\Big\} 
 + \omega_{g} \ , \ \ \ \ \label{ome}
\end{eqnarray}
where the energy and the constituent quark mass are defined by $E_p = \sqrt{p^2+M^2}$ and $M = m_q - G\sigma$, respectively, 
with the current quark mass $m_q$, the coupling constant $G$ and the chiral condensate $\sigma$. 
The Polyakov loop $L$ is defined by 
\begin{eqnarray}
L(\vec{x}) &=& {\cal P} \exp \brac{ i \int_{0}^{1/T} dx_4 \, A_4 (\vec{x},x_4) } \ ,
\end{eqnarray}
where ${\cal P}$ stands for the path ordering and $A_4=iA_0$ is the $SU(N_c)$ temporal-gauge field in Euclidian space. 
Moreover, we express the polynomial Polyakov-loop potential as the gauge-field contribution of the effective potential, 
\begin{eqnarray}
\omega_{g}(T,\mu) &=& T^4 \brac{-\frac{b_2(T)}{2} \ell\bar\ell 
                                                                  -\frac{b_3}{6}\br{\ell ^3+\bar\ell ^3} 
                                                                  +\frac{b_4}{4}\br{\ell \bar\ell}^2 } , \ \ \ \ \label{omeg}
\end{eqnarray}
where $\ell$ and $\bar \ell$ are the thermal expectation values of the color trace of the Polyakov loop and its conjugate, 
\begin{eqnarray}
\ell (\vec{x}) \equiv \frac{1}{N_c} \left \langle \Tr_c L(\vec{x})  \right \rangle \ , \ \ \
\bar\ell (\vec{x}) \equiv \frac{1}{N_c} \left \langle \Tr_c L^{\dag}(\vec{x})  \right \rangle \ . \ \ 
\end{eqnarray}
Note that $\Tr_c L$ and $\Tr_c L^{\dag}$ are generally complex in $SU(N_c)$ for $N_c\ge 3$. 
We choose the parameters in Eq.~\eqref{omeg} as in Ref.~\cite{Skokov:2010uh}: 
\begin{eqnarray}
b_2(T) &=& a_0 + a_1\br{\frac{T_0}{T}} + a_2\br{\frac{T_0}{T}}^2 + a_3\br{\frac{T_0}{T}}^3 , \ \ \ \
\end{eqnarray}
$a_0=6.75$, $a_1=-1.95$, $a_2=2.625$, $a_3=-7.44$, $b_3=0.75$, $b_4=7.5$ and $T_0 = 270$~[MeV]. 

In case of $N_c=3$, Polyakov loops are represented as 
$L={\rm diag}(e^{i\varphi_1},e^{i\varphi_2},e^{-i(\varphi_1+\varphi_2)})$ 
under the Polyakov gauge. 
Therefore, we can rewrite the color traces in Eq.~\eqref{ome} as follows, 
\begin{eqnarray}
&&\Tr_c  \ln\brac{1+ L e^{-\frac{E_p-\mu}{T}}} \non\\
 &=& \ln \brac{1+\Tr_c{L}  e^{-\frac{E_p-\mu}{T}} + \Tr_c{L^{\dag}}  e^{-\frac{2\br{E_p-\mu}}{T}}+e^{-\frac{3\br{E_p-\mu}}{T}}} \non\\
 &\to& \ln \brac{1+3\ell  e^{-\frac{E_p-\mu}{T}} +3 \bar\ell  e^{-\frac{2\br{E_p-\mu}}{T}}+e^{-\frac{3\br{E_p-\mu}}{T}}} \ , \\
 && \Tr_c  \ln\brac{1+ L^{\dag} e^{-\frac{E_p+\mu}{T}}} \non\\
&=& \ln \brac{1+\Tr_c{L^{\dag}}  e^{-\frac{E_p+\mu}{T}} +\Tr_c{L}  e^{-\frac{2\br{E_p+\mu}}{T}}+e^{-\frac{3\br{E_p+\mu}}{T}}}  \non \\
&\to& \ln \brac{1+3\bar\ell  e^{-\frac{E_p+\mu}{T}} +3 \ell  e^{-\frac{2\br{E_p+\mu}}{T}}+e^{-\frac{3\br{E_p+\mu}}{T}}} \ ,
\end{eqnarray}
where we replace $\Tr_c L$ and $\Tr_c L^\dag$ to $\ell$ and $\bar\ell$ 
with the mean field approximation in the third lines of each equation. 
The values of $\ell$, $\bar\ell$ and $\sigma$ are obtained from a solution of the gap equations 
which comes from the three stationary conditions: 
\begin{eqnarray}
\frac{\partial \omega}{\partial \sigma} =
\frac{\partial \omega}{\partial \ell } =
\frac{\partial \omega}{\partial \bar\ell } = 0 \ .
\end{eqnarray}

\subsection{The PNJL model at the pure imaginary chemical potential}

In this paper, we compute $n_{qI}$ in Eq.~\eqref{mulsin} in the PNJL model. 
Practically, it is convenient to evaluate $n_{qI}$ numerically with the difference approximation such as  
\begin{eqnarray}
n_{qI}(\mu_I) &=&  \frac{1}{T} \frac{\partial \omega}{\partial \br{\mu_I/T}} \non\\
 &\approx& \frac{ \omega\br{\mu_{I}/T+\delta\br{\mu_{I}/T}} - \omega\br{\mu_{I}/T-\delta\br{\mu_{I}/T}} }{2T\delta\br{\mu_{I}/T}} , \non \\
 && \label{nq}
\end{eqnarray}
where we use $\delta\br{\mu_{I}/T}=10^{-18}$. 
The calculations of $n_{qI}$ are carried out with 128 significant digits in decimal notation 
by using a multiple-precision arithmetic package, FMLIB~\cite{FMLIB}. 

In the pure imaginary chemical potential, $\ell$ and $\bar \ell$ are complex 
but $\bar\ell$ is the same as the complex conjugate of $\ell(\equiv \ell_r e^{i\ell_\phi})$, 
$\bar\ell=\ell^{\dag}=\ell_r e^{-i\ell_\phi}$, where $\ell_r$ and $\ell_\phi$ are real. 
Therefore, $\omega(\mu_I/T)$ is obtained from the three stationary conditions: 
\begin{eqnarray}
\frac{\partial \omega}{\partial \sigma} =
\frac{\partial \omega}{\partial \ell_r } =
\frac{\partial \omega}{\partial \ell_\phi } = 0 \ .
\end{eqnarray}
The conditions correspond to the three gap equations as follows: 
\begin{eqnarray}
M &=&  m_q + \frac{3 N_f GM}{\pi^2} \int_{0}^{\Lambda} dp  \frac{p^2}{E_p} \Bigg[1 \non \\
    &&     - \frac{ \ell  e^{-\frac{E_p-i\mu_I}{T}} +2 \ell^{\ast}  e^{-\frac{2\br{E_p-i\mu_I}}{T}}+e^{-\frac{3\br{E_p-i\mu_I}}{T}} }
                       {1+3\ell  e^{-\frac{E_p-i\mu_I}{T}} +3 \ell^{\ast}  e^{-\frac{2\br{E_p-i\mu_I}}{T}}+e^{-\frac{3\br{E_p-i\mu_I}}{T}} } \non\\
    &&     - \frac{ \ell^{\ast}  e^{-\frac{E_p+i\mu_I}{T}} +2 \ell  e^{-\frac{2\br{E_p+i\mu_I}}{T}}+e^{-\frac{3\br{E_p+i\mu_I}}{T}} }
                       {1+3\ell^{\ast}  e^{-\frac{E_p+i\mu_I}{T}} +3 \ell  e^{-\frac{2\br{E_p+i\mu_I}}{T}}+e^{-\frac{3\br{E_p+i\mu_I}}{T}}} \Bigg] , \non\\
    && \label{Mgap}
\end{eqnarray}
\begin{eqnarray}
 \ell_r  
 &=&\frac{1}{b_2(T)}\Bigg[ -b_3\ell_r ^2 \cos\br{3\ell_\phi}   +b_4 \ell_r^3 
  -\frac{3N_f}{\pi^2 T^3} \int_{0}^{\Lambda} dp \, p^2 \Bigg\{ \non\\
 &&                                \frac{  e^{i\ell_\phi} e^{-\frac{E_p-i\mu_I}{T}}  + e^{-i\ell_\phi}e^{-\frac{2\br{E_p-i\mu_I}}{T}} }
                                        {1+3\ell  e^{-\frac{E_p-i\mu_I}{T}} +3 \ell^{\ast}  e^{-\frac{2\br{E_p-i\mu_I}}{T}}+e^{-\frac{3\br{E_p-i\mu_I}}{T}} } \non\\
 && \!\!\!\!
                                + \frac{ e^{-i\ell_\phi} e^{-\frac{E_p+i\mu_I}{T}} + e^{i\ell_\phi} e^{-\frac{2\br{E_p+i\mu_I}}{T}} }
                                          {1+3\ell^{\ast}  e^{-\frac{E_p+i\mu_I}{T}} +3 \ell  e^{-\frac{2\br{E_p+i\mu_I}}{T}}+e^{-\frac{3\br{E_p+i\mu_I}}{T}}} \Bigg\} \Bigg] , \non\\
   && \label{rgap}
\end{eqnarray}
\begin{eqnarray}
  \sin \br{\ell_\phi} &=& \frac{4}{3}\sin^3 \br{\ell_\phi} + \frac{i N_f}{\pi^2 b_3\ell_r ^3T^3}  \int_{0}^{\Lambda} dp \, p^2  \non \\
&& \!\!\!\!\!\!\!\!\!\!                      \Bigg[ \frac{  \ell e^{-\frac{E_p-i\mu_I}{T}}  - \ell^{\ast} e^{-\frac{2\br{E_p-i\mu_I}}{T}} }
                                  {1+3\ell  e^{-\frac{E_p-i\mu_I}{T}} +3 \ell^{\ast}  e^{-\frac{2\br{E_p-i\mu_I}}{T}}+e^{-\frac{3\br{E_p-i\mu_I}}{T}} } \non\\
 && \!\!\!\!\!\!\!\!\!\!\!\!\!\!                      + \frac{  - \ell^{\ast} e^{-\frac{E_p+i\mu_I}{T}} +  \ell e^{-\frac{2\br{E_p+i\mu_I}}{T}} }
                                          {1+3\ell^{\ast}  e^{-\frac{E_p+i\mu_I}{T}} +3 \ell  e^{-\frac{2\br{E_p+i\mu_I}}{T}}+e^{-\frac{3\br{E_p+i\mu_I}}{T}}} \Bigg] . \non\\
 && \label{pgap}
\end{eqnarray}
Note that $M$ is real in the pure imaginary chemical potential. 
We take 
$N_f=2$, $m_q=5.5$~[MeV], $G=0.214$~[fm$^{2}$] and the tree-momentum cutoff $\Lambda=631$~[MeV], respectively, 
which are fixed to reproduce 
the pion decay constant $f_\pi=93$~[MeV] and the constituent quark mass $M=335$~[MeV] 
in the mean field approximation. 
  
\begin{figure}
\begin{center}
\includegraphics[scale=0.40]{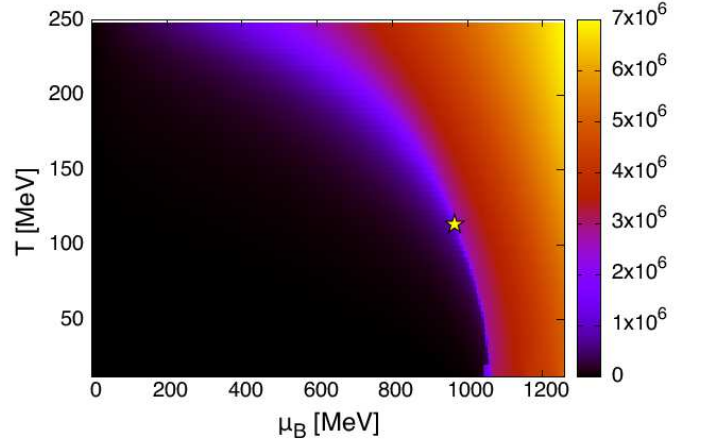}
\caption{ (color online). 
The temperature and chemical potential dependences of the number density in the PNJL model. 
The star is the critical end point (CEP) $(T^{\rm CEP},\mu^{\rm CEP}_B)\simeq(114,965)$~[MeV]. 
}
\label{nq_r}
\end{center}
\end{figure}

\section{\label{simu}Numerical results}

\subsection{\label{nqr} Exact results in the PNJL model}

Figure~\ref{nq_r} shows the exact results of the real baryon number density $n_B=n_q/3$ depending on temperature and baryon chemical potential $(\mu_B=3\mu)$ in the PNJL model. 
The critical end point (CEP): $(T^{\rm CEP},\mu^{\rm CEP}_B)\simeq(114,965)$~[MeV] is represented as a star in Fig.~\ref{nq_r}. 
These results are close to the previously obtained results~\cite{Fukushima:2003fw}, 
and will be compared with the results in the following subsections.

\begin{figure}
\begin{center}
\includegraphics[scale=0.425]{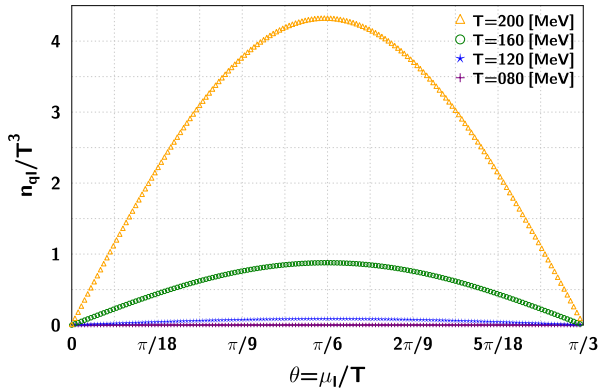}
\caption{ (color online). 
The $\theta$ dependence of the imaginary number density in the PNJL model. 
}
\label{thetadep}
\end{center}
\end{figure}

\begin{table}
\caption{
The coefficients $f_{3k}$ from the data of $n_{qI}/T^3$ for each temperature.}
\begin{tabular}{c|cccc}\hline \hline
 $T$\ [MeV] & $f_3$ & $f_6$ & $f_{9}$ & $f_{12}$  \\\hline 
200 & $2.2\times10^{-2}$ & $1.7\times10^{-4\ }$ & $1.9\times10^{-6\ }$ & $2.2\times10^{-8\ }$   \\
160 & $5.5\times10^{-3}$ & $9.0\times10^{-6\ }$ & $2.1\times10^{-8\ }$ & $5.8\times10^{-11}$   \\
120 & $7.2\times10^{-4}$ & $9.9\times10^{-8\ }$ & $2.0\times10^{-11}$ & $4.7\times10^{-15}$    \\
\ 80 & $1.4\times10^{-5}$ & $1.7\times10^{-11}$ & $3.0\times10^{-17}$ & ---   \\ \hline \hline
\end{tabular}
\label{table_coef}
\end{table}

\subsection{\label{nqi} Imaginary number density in the PNJL model}

We evaluate the imaginary number density $n_{qI}$ at the pure imaginary chemical potential 
from Eq.~\eqref{nq}. 
The momentum integrations in Eqs.~\eqref{Mgap}, \eqref{rgap} and \eqref{pgap} are calculated with the Gaussian quadrature method. 
Figure~\ref{thetadep} shows the $\theta$ (=$\mu_I/T$) dependence of the imaginary number density. 
$n_{qI}/T^3$ are calculated at 161 values of $\mu_{I}$ for various temperatures. 
The PNJL model has the $Z_3$ symmetry and an anti-symmetry such as 
$n_{qI}(\theta)=n_{qI}(\theta+2\pi/3)$ and $n_{qI}(\theta)=-n_{qI}(-\theta)$. 
Therefore, we only show the region $0 \le \theta \le \pi/3$ in Fig.~\ref{thetadep}. 
From Fig.~\ref{thetadep}, we find that 
$n_{qI}$ is well approximated by the Fourier series 
\begin{eqnarray}
 \frac{n_{qI}}{T^3}(\theta) &=& \sum_{k=1}^{N_{{\sin}}} f_{3k} \sin(3k\theta) \ , 
 \label{mulsin2}
\end{eqnarray}
which is used instead of Eq.~\eqref{mulsin} since $f_k$ for $\mod(k,3)\neq 0$ are zero due to the $Z_3$ symmetry. 
Since we are interested in the confinement phase of QCD here, 
the $Z_3$ symmetric feature in Eq.~(\ref{mulsin2}) remains intact. 
The obtained coefficients $f_{3k}$ are listed in Table~\ref{table_coef}.

\begin{figure}
\begin{center}
\includegraphics[scale=0.44]{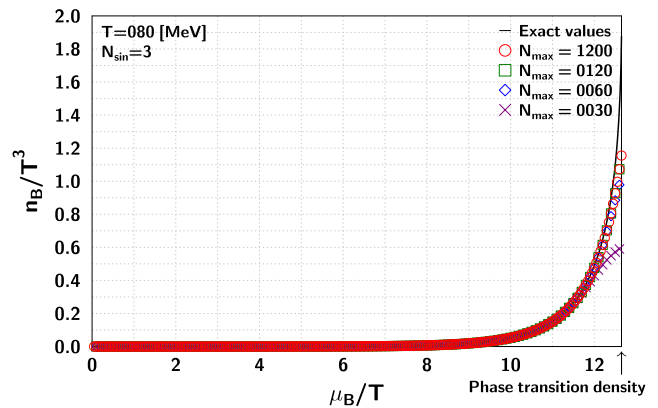}
\caption{ (color online). 
The $N_{\rm max}$ dependence of the number density in the PNJL model. 
The solid line is the exact number density calculated at the real chemical potential. 
The other symbols are the number densities obtained from the canonical approach for several $N_{\rm max}$. 
}
\label{nmaxdep}
\end{center}
\end{figure}

\begin{figure*}
\begin{center}
\includegraphics[scale=0.29701]{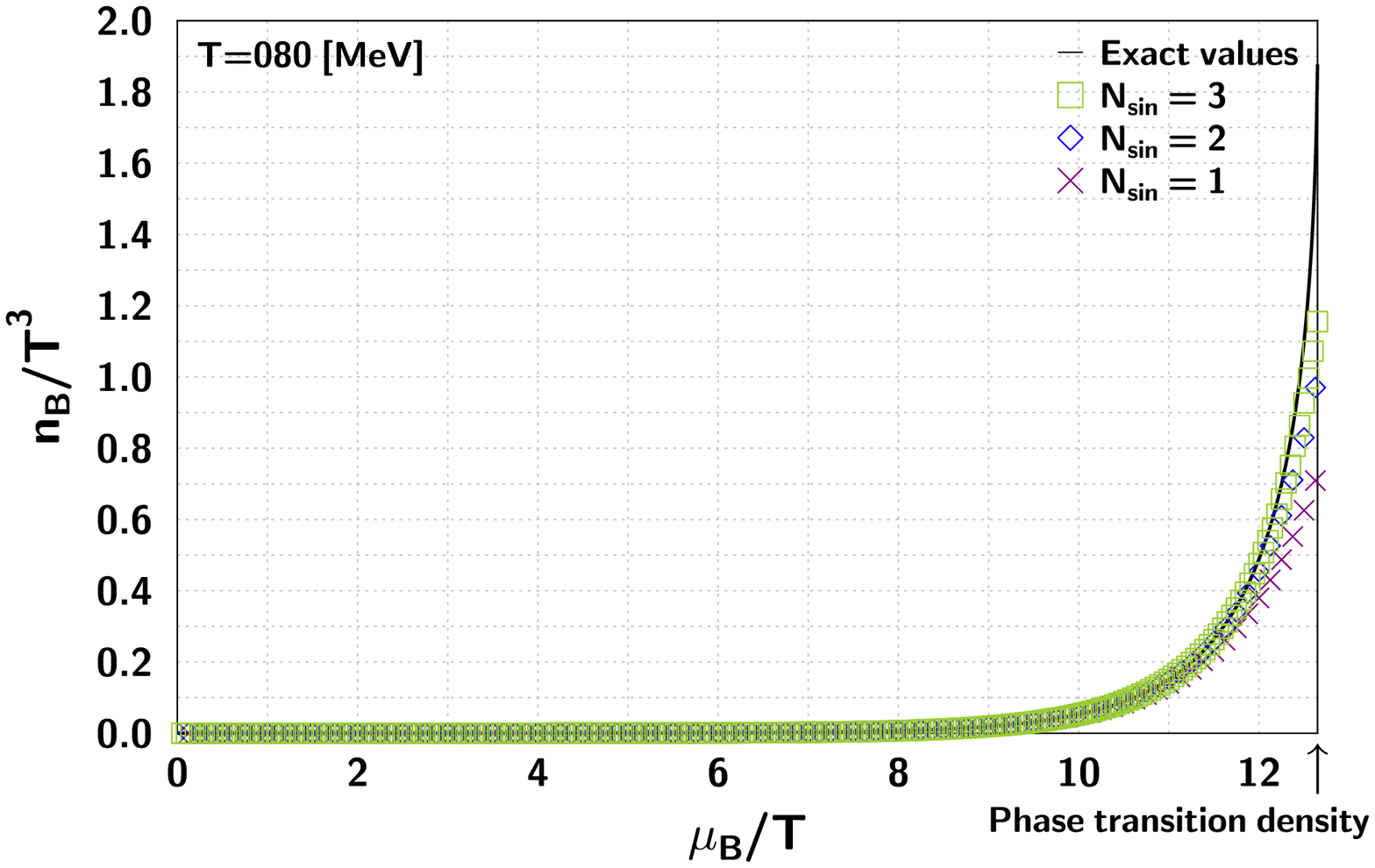}
\includegraphics[scale=0.29701]{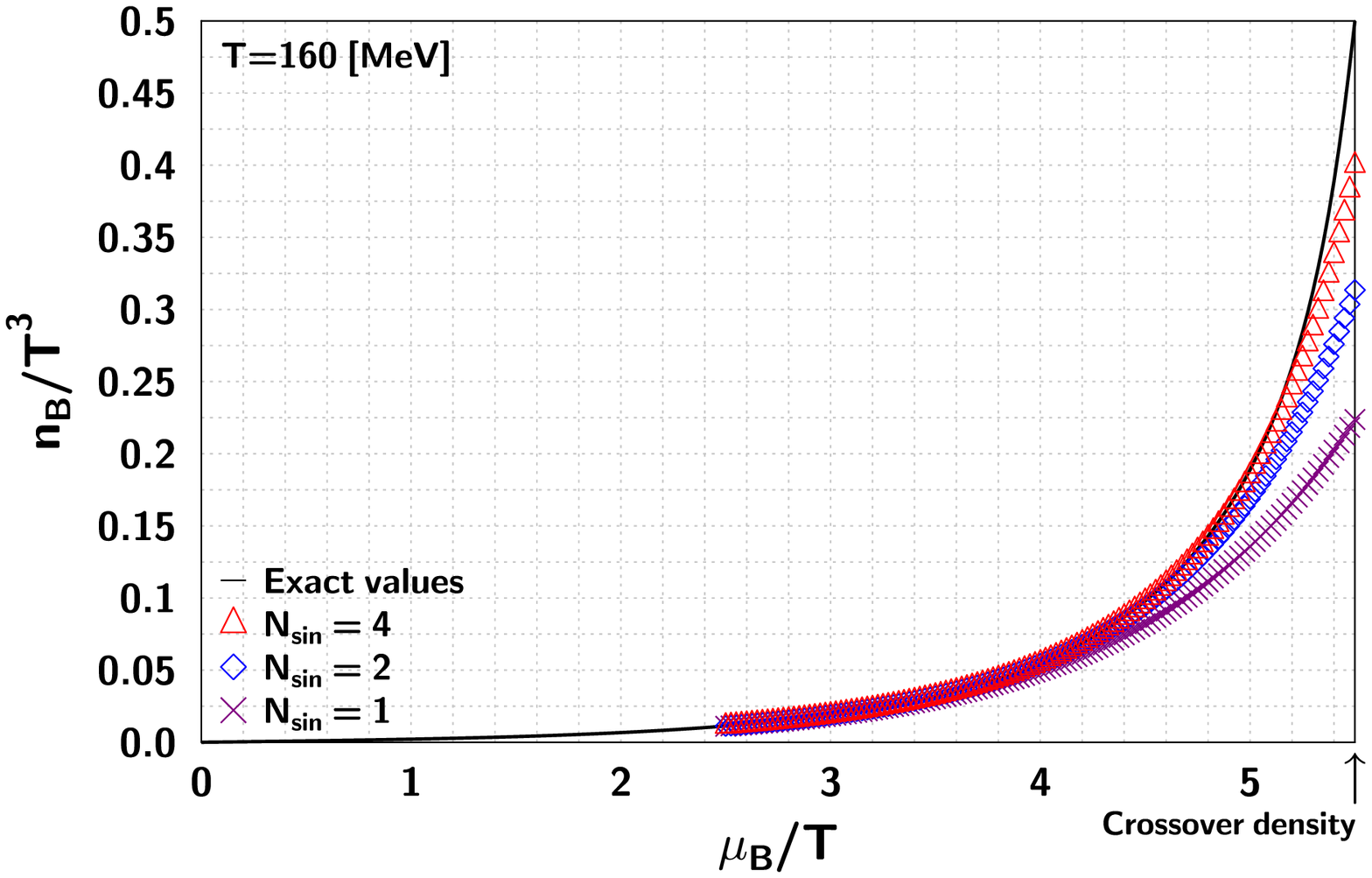}
\includegraphics[scale=0.29701]{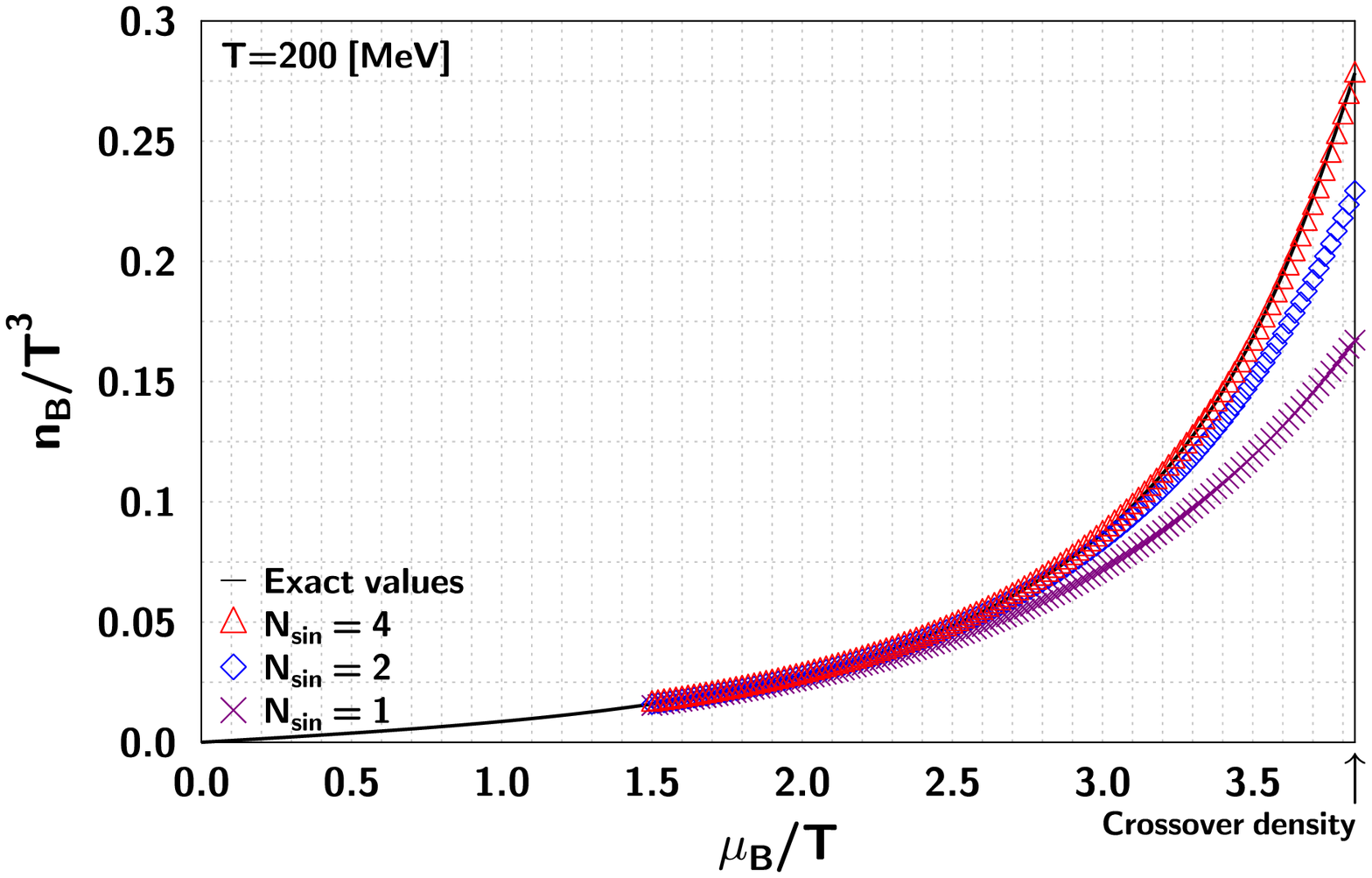}
\caption{ (color online). 
The $N_{\sin}$ dependence of $n_B/T^3$ in the PNJL model. 
The solid lines are the exact number densities calculated at the real chemical potential. 
}
\label{nsindep1}
\end{center}
\end{figure*}

\begin{figure*}
\begin{center}
\includegraphics[scale=0.29701]{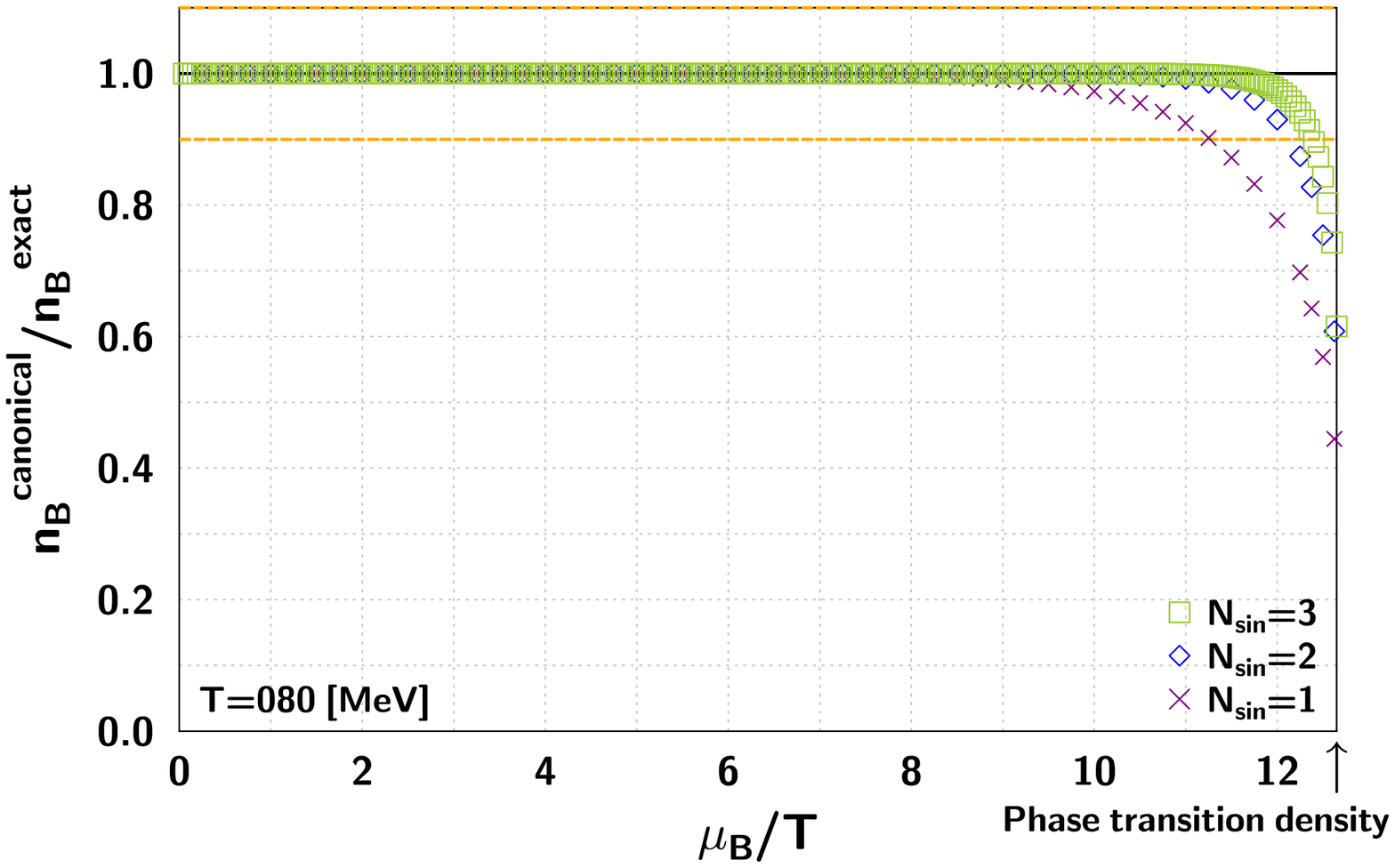}
\includegraphics[scale=0.29701]{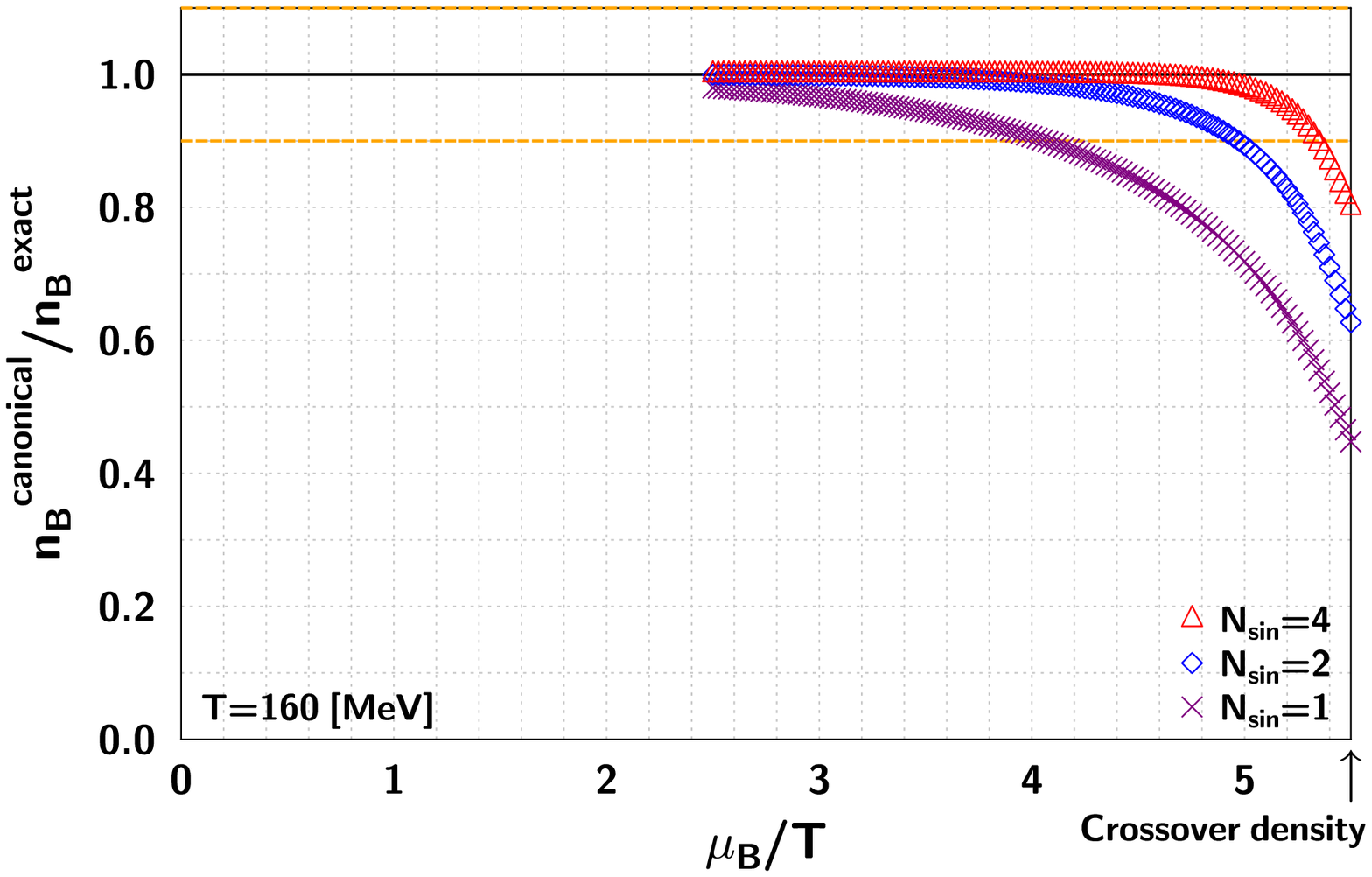}
\includegraphics[scale=0.29701]{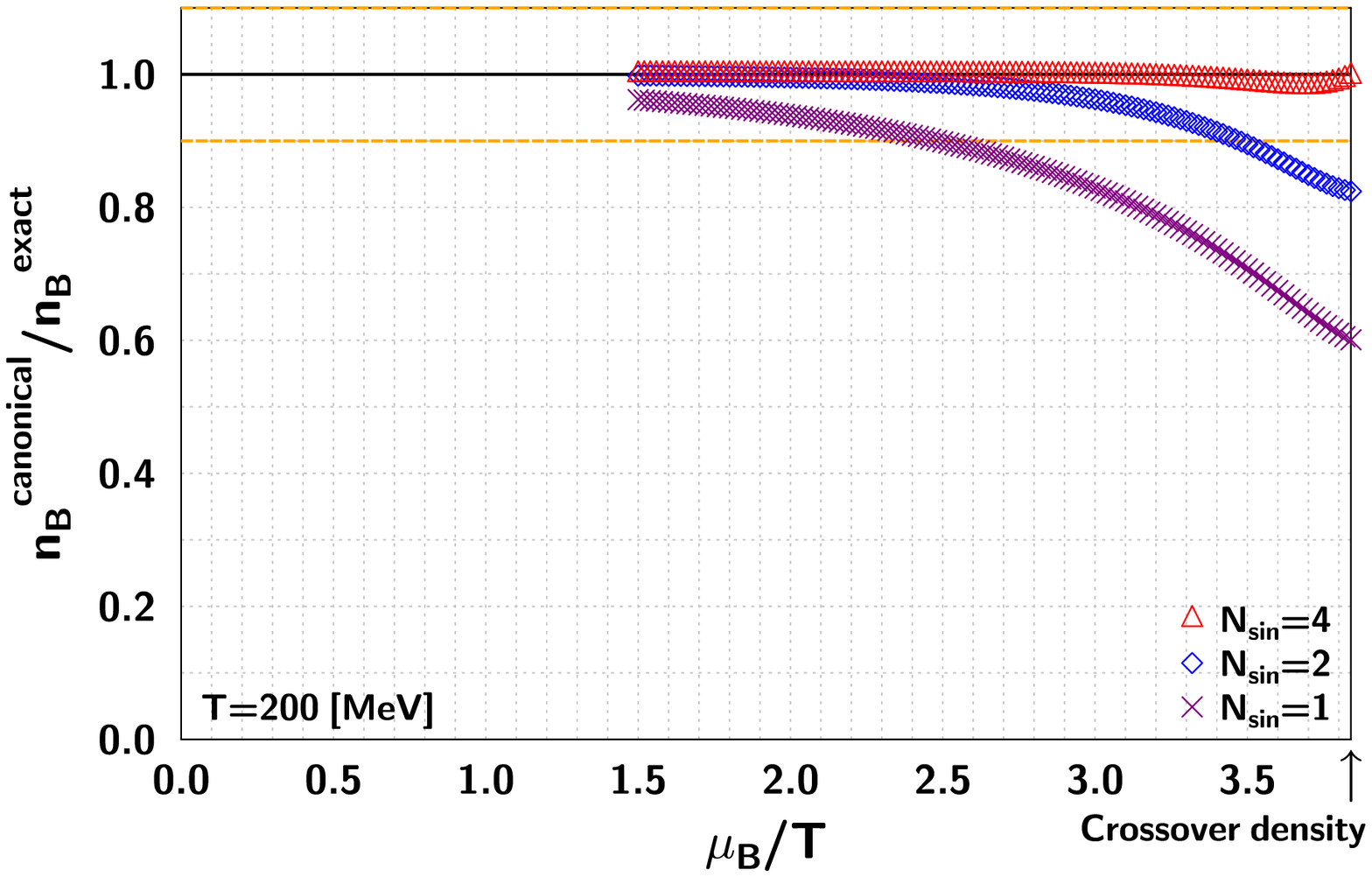}
\caption{ (color online). 
The $N_{\sin}$ dependence of $n_B^{\rm canonical}/n_B^{\rm exact}$ in the PNJL model. 
$n_B^{\rm canonical}$ is the number density obtained from the canonical approach and 
$n_B^{\rm exact}$ is the exact number density calculated at the real chemical potential. 
}
\label{nsindep2}
\end{center}
\end{figure*}

\subsection{\label{nmax}$N_{\rm max}$ dependence of the number density in the PNJL model}

Next, we calculate the grand canonical partition function at pure imaginary chemical potential 
with the integration method in Eq.~\eqref{GCint}. 
Here, the finite volume effect is included as the coefficient $V$ in Eq.~\eqref{GCint}, 
although the imaginary number densities and $f_{3k}$ in Eq.~\eqref{mulsin2} are computed by the formula for the infinite volume. 
In this paper, since we study the $N_{\rm max}$ and $N_{\sin}$ dependences of the canonical approach, 
we use $V = (6\ [{\rm fm}])^{3}$ to minimize the finite $V$ effect, 
which is justified in comparison with the argument of Ref.~\cite{Xu:2019gia}, 
where $V \sim (5\ [{\rm fm}])^{3}$ is shown to be sufficiently large. 

By performing Fourier transforms in Eq.~\eqref{ZC} with 8,192 significant digits in decimal notation, 
we obtain the canonical partition functions. 
Finally, we can reconstruct the grand canonical partition function, 
\begin{eqnarray}
Z _{\rm GC} (\mu,T,V) 
&=& \sum_{n=-N_{\rm max}}^{N_{\rm max}} Z_C(n,T,V) \xi^n \ , 
\label{fuga_exp_nmax}
\end{eqnarray}
where 
$N_{\rm max}$ is a maximum value of fluctuation of the net quark number in the system. 
We should take $N_{\rm max}$ to an infinite limit theoretically, 
but a numerical constraint makes $N_{\rm max}$ finite.

In Fig.~\ref{nmaxdep}, 
we present the $N_{\rm max}$ dependence of the baryon number density $n_B$ obtained from the canonical approach at $T=80$~[MeV]. 
The solid line is the exact number density calculated at the real chemical potential. 
The figure only shows up to the exact phase transition density 
because the Fourier transforms in the canonical approach are ineffective over the phase transition point. 
From Fig.~\ref{nmaxdep}, 
we find that the behavior of the number density converges for $N_{\rm max}=120$ and larger. 
Note that the difference between $n_B$ calculated from the canonical approach and the exact values near the phase transition density 
comes from the finite $N_{\sin}$ effect, which we discuss in the next subsection. 
Now we can understand the converging behavior of $n_B$ by comparing 
$N_{\rm max}/(3V)=120/(3\times6^3)\sim0.19$~[fm$^{-3}$] with the normal nuclear matter density 0.17~[fm$^{-3}$]. 
It is reasonable to expect that the fluctuations of the number density are in the same order of the nuclear matter density 
in the region of the chemical potential and temperature that we are looking at now.

\subsection{\label{nsin}$N_{\sin}$ dependence of the number density in the PNJL model}

In this subsection, 
we discuss the $N_{\sin}$ dependence by using $N_{\rm max}=1200$ to suppress possible uncertainties due to finite $N_{\rm max}$. 
In Fig.~\ref{nsindep1}, we show the $N_{\sin}$ dependence of the baryon number density at $T=80,$ 160 and 200~[MeV]. 
The solid lines are the exact number densities $n_B^{\rm exact}$ calculated at the real chemical potential. 
The symbols represent the number densities obtained from the canonical approach, $n_B^{\rm canonical}$. 
As $N_{\sin}$ increases, the difference between $n_B^{\rm exact}$ and $n_B^{\rm canonical}$ becomes small. 

\begin{figure}
\begin{center}
\includegraphics[scale=0.72]{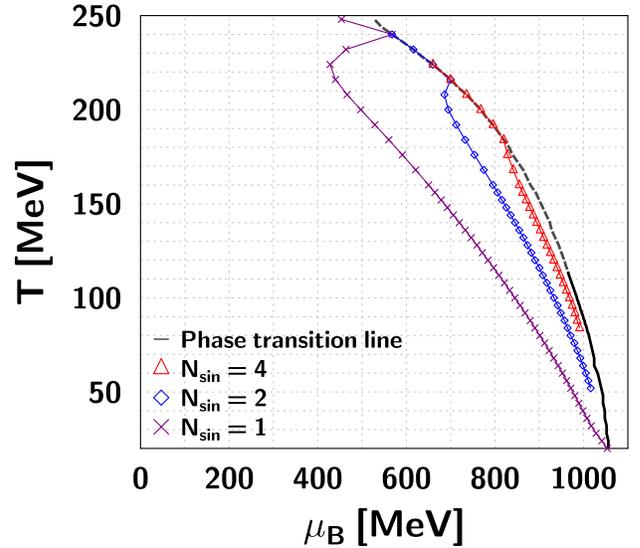}
\caption{ (color online). 
The boundaries of the effective regions of the canonical approach for $N_{\sin}$ in the PNJL model. 
The black solid and dashed lines represent the first-order phase transition and crossover lines, respectively. 
The points of making a 10\% difference between the exact $n_B$ and the results from the canonical approach are plotted. 
We plot the symbols on the crossover line when the difference is less than 10\% in the confinement phase. 
}
\label{nsindep3}
\end{center}
\end{figure}

In Fig.~\ref{nsindep2}, we show the $N_{\sin}$ dependence of the ratio of $n_B^{\rm canonical}$ to $n_B^{\rm exact}$ 
at $T=80,$ 160 and 200~[MeV]. 
The solid and dashed lines represent the exact value ($n_B^{\rm canonical}/n_B^{\rm exact}=1.0$) 
and the 10\% difference values ($n_B^{\rm canonical}/n_B^{\rm exact}=0.9$ and 1.1), respectively. 
In this paper, we define the density region having a difference of less than 10\% as the effective region of the canonical approach. 
For $N_{\sin}=1$ at $T=80$, 160 and 200~[MeV], 
the boundaries of the effective region of the canonical approach 
appear at the 89\%, 74\% and 65\% of the phase transition or crossover densities, respectively. 
It turns out that 
as the temperature decreases, 
the Fourier series approximation with $N_{\sin}=1$ becomes better. 
For $N_{\sin}=3$ at $T=80$~[MeV] and $N_{\sin}=4$ at $T=160$~[MeV], 
we can reconstruct the exact baryon number density from the canonical approach 
until {(97 -- 98) \% of the phase transition or crossover density within the 10\% difference. 
Moreover, for $N_{\sin}=4$ at $T=200$~[MeV], 
$n_B^{\rm canonical}$ only appears the difference less than 1.8\% from the exact value until the crossover density.

In Fig.~\ref{nsindep3}, we plot the symbols on the boundaries of the effective region of the canonical approach 
for each $N_{\sin}$ and temperature. 
In the left regions of the boundaries, we can discuss $n_B$ within the 10\% difference from the canonical approach. 
When the difference is less than 10\% in the confinement phase, we plot the symbols on the crossover density 
as the high-density limits of the effective region, 
such as at $T=(184 - 224)$~[MeV] for $N_{\sin}=4$. 
The reason is that there is no crossover or phase transition structure 
in the Fourier series approximation with finite $N_{\sin}$ since the function is analytic. 
From Fig.~\ref{nsindep3}, 
we find that most of the confinement phase can be reliably studied by the canonical approach with $N_{\sin}=4$. 
Furthermore, 
for $T < T^\mathrm{CEP}$ and $\mu_B < 900$~[MeV], 
$N_{\sin}=1$ or 2 is enough to reconstruct the exact number density from the canonical approach. 
The results suggest that the application of the canonical approach to the lattice QCD is useful, especially in the confinement phase.

\subsection{\label{njl_pnjl}Comparison with the NJL and PNJL models}

At the end of this section, 
we consider the model dependence by comparing the results of the PNJL model with those of the NJL one. 
In the NJL model, 
we obtain the coefficients $f_{k}$ from 161 values of data of $n_{qI}/T^3$ such as Table~\ref{table_coefNJL}. 
Here, we use not Eq.~\eqref{mulsin2} but Eq.~\eqref{mulsin} since the NJL model does not have the $Z_3$ symmetry. 
As it was done in the PNJL model, 
we set $V$ in Eq.~\eqref{integration} to $(6\ [{\rm fm}])^{3}$ 
and reconstruct the grand canonical partition function 
by performing the Fourier transforms with 8,192 significant digits in decimal notation. 

\begin{table}
\caption{
The coefficients $f_{k}$ from the data of $n_{qI}/T^3$ for each temperature in the NJL model.}
\begin{tabular}{c|cccc}\hline \hline
 $T$\ [MeV] & $f_1$ & $f_2$ & $f_{3}$ & $f_{4}$  \\\hline 
79 & $2.7\times10^{-1}$ & $2.3\times10^{-3}$ & $2.9\times10^{-5\ }$ & $4.2\times10^{-7\ }$   \\
49 & $3.7\times10^{-2}$ & $1.8\times10^{-5}$ & $1.3\times10^{-8\ }$ & $1.1\times10^{-11}$   \\
29 & $6.5\times10^{-4}$ & $1.9\times10^{-9}$ & $7.9\times10^{-15}$ & $3.9\times10^{-20}$   \\ \hline \hline
\end{tabular}
\label{table_coefNJL}
\end{table}

Figure~\ref{nmaxdepNJL} shows the $N_{\max}$ dependence of the baryon number density at $T=49$~[MeV] in the NJL model. 
The solid line is the exact number density calculated at the real chemical potential. 
We find that the behavior of the number density converges for $N_{\rm max}=120$ and larger, 
which is the same as the result of the PNJL model. 
In the following discussion for the NJL model, we use $N_{\rm max}=400$. 

In Fig.~\ref{nsindep1NJL}, we show the $N_{\sin}$ dependence of the number density at $T=29$, 49 and 79~[MeV] in the NJL model. 
The solid lines are the exact number densities $n_B^{\rm exact}$ calculated at the real chemical potential. 
The symbols represent the number densities obtained from the canonical approach, $n_B^{\rm canonical}$. 
As $N_{\sin}$ increases, the difference between $n_B^{\rm exact}$ and $n_B^{\rm canonical}$ becomes small. 

In Fig.~\ref{nsindep2NJL}, we show the $N_{\sin}$ dependence of the ratio of $n_B^{\rm canonical}$ to $n_B^{\rm exact}$ 
in the NJL model. 
For $N_{\sin}=4$ at $T=29$, 49 and 79~[MeV], 
we can reconstruct the exact baryon number density from the canonical approach 
until 99\%, 97\% and 96\% of the phase transition or crossover density within the 10\% difference, respectively. 

In Fig.~\ref{nsindep3NJL}, we plot the symbols on the high-density limits of the effective region of the canonical approach 
for each $N_{\sin}$ and temperature in the NJL model. 
We find that the effective region of the canonical approach for $N_{\sin}=4$ can cover in most of the confinement phase. 
For $T\lesssim49\sim T^\mathrm{CEP}$~[MeV] and $\mu_B\lesssim900$~[MeV], 
$N_{\sin}=1$ or 2 is enough to reconstruct the exact number density from the canonical approach. 
The results have universality for at least the NJL and PNJL  models.

\begin{figure}
\begin{center}
\includegraphics[scale=0.44]{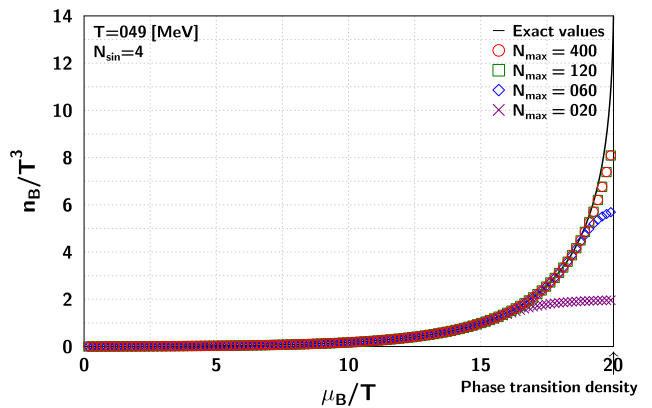}
\caption{ (color online). 
The $N_{\rm max}$ dependence of the number density in the NJL model. 
The solid line is the exact number density calculated at the real chemical potential. 
The other symbols are the number densities obtained from the canonical approach for several $N_{\rm max}$. 
}
\label{nmaxdepNJL}
\end{center}
\end{figure}

\begin{figure*}
\begin{center}
\includegraphics[scale=0.29701]{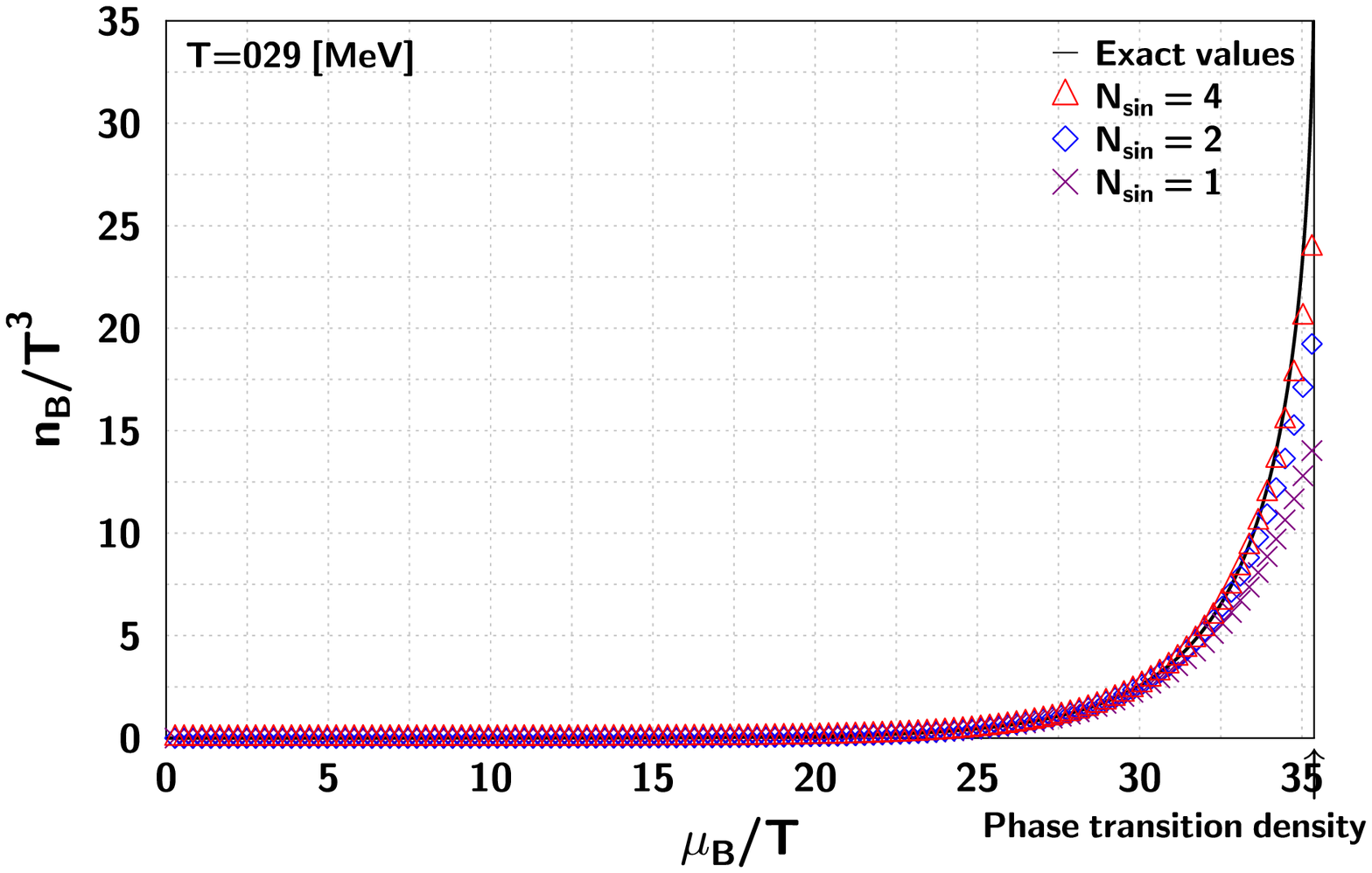}
\includegraphics[scale=0.29701]{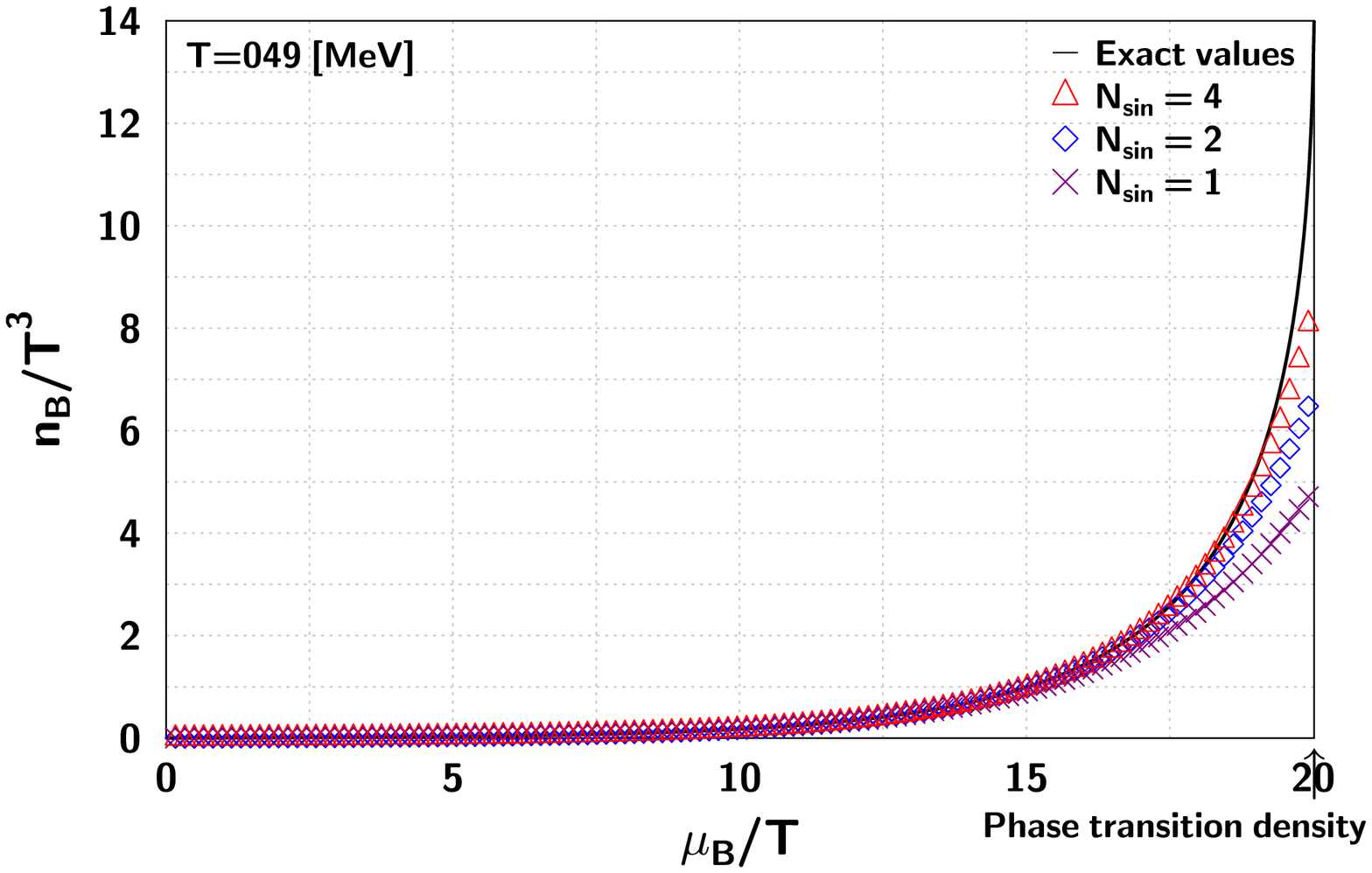}
\includegraphics[scale=0.29701]{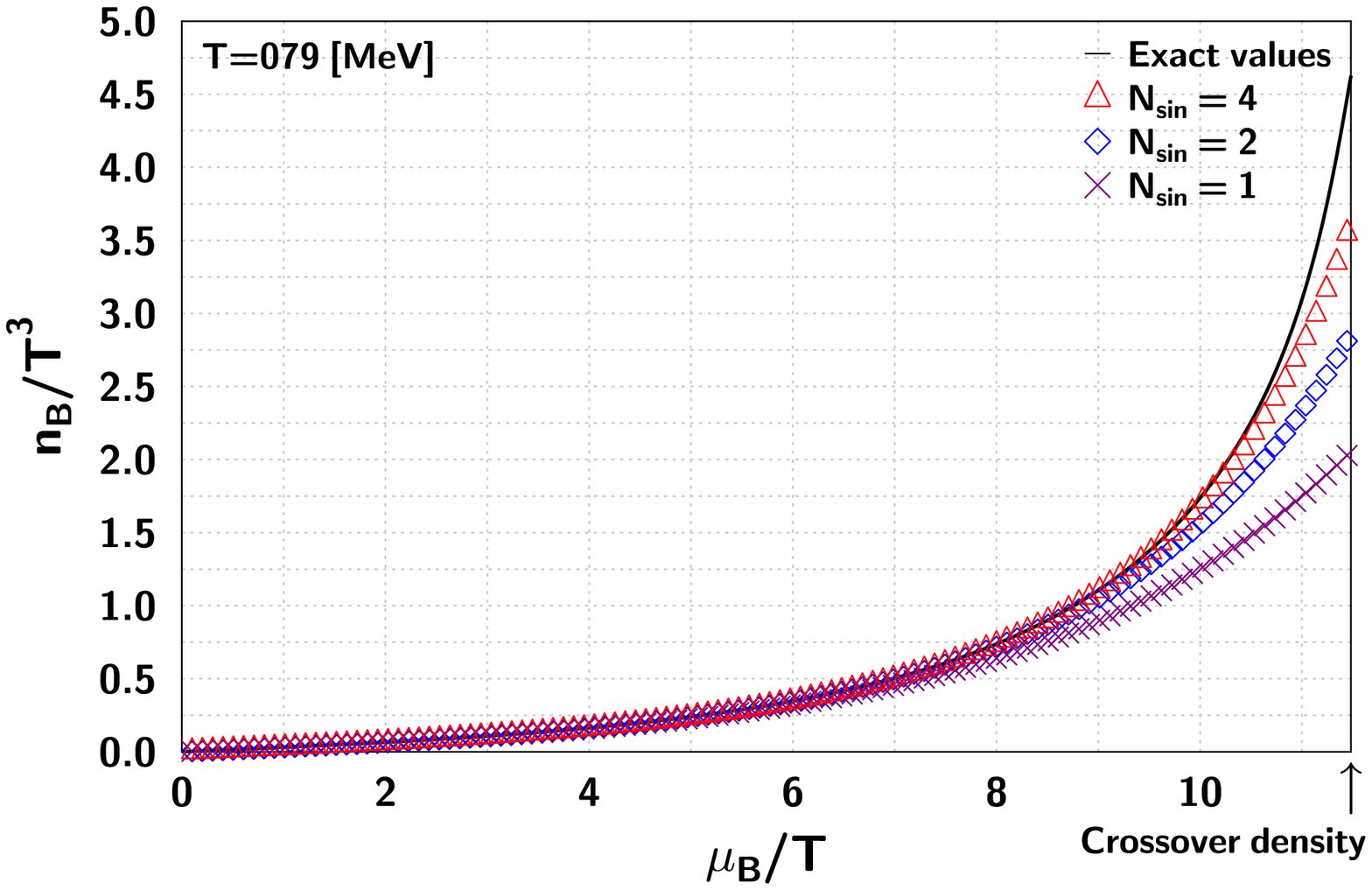}
\caption{ (color online). 
The $N_{\sin}$ dependence of $n_B/T^3$ in the NJL model. 
The solid lines are the exact number densities calculated at the real chemical potential. 
}
\label{nsindep1NJL}
\end{center}
\end{figure*}

\begin{figure*}
\begin{center}
\includegraphics[scale=0.29701]{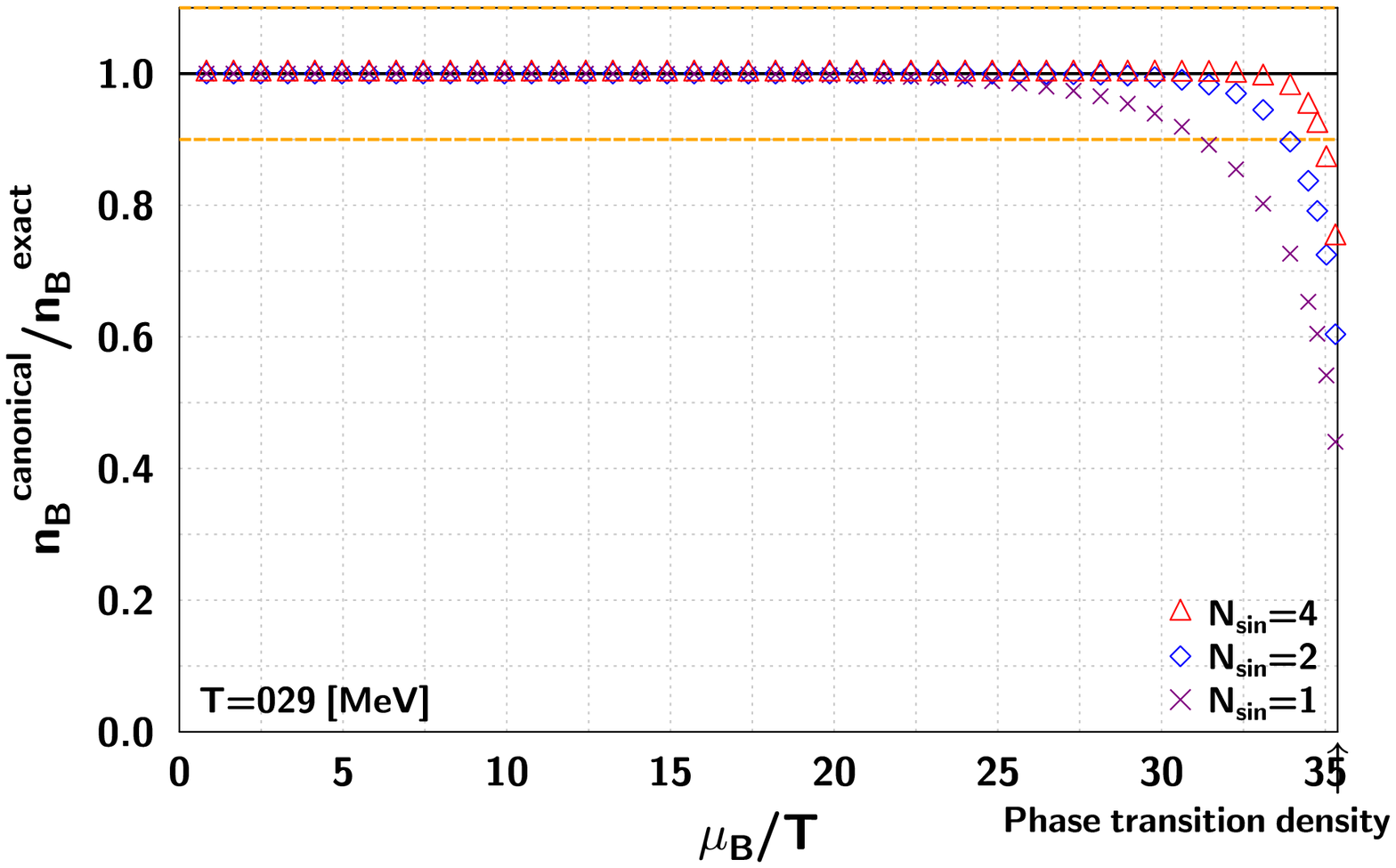}
\includegraphics[scale=0.29701]{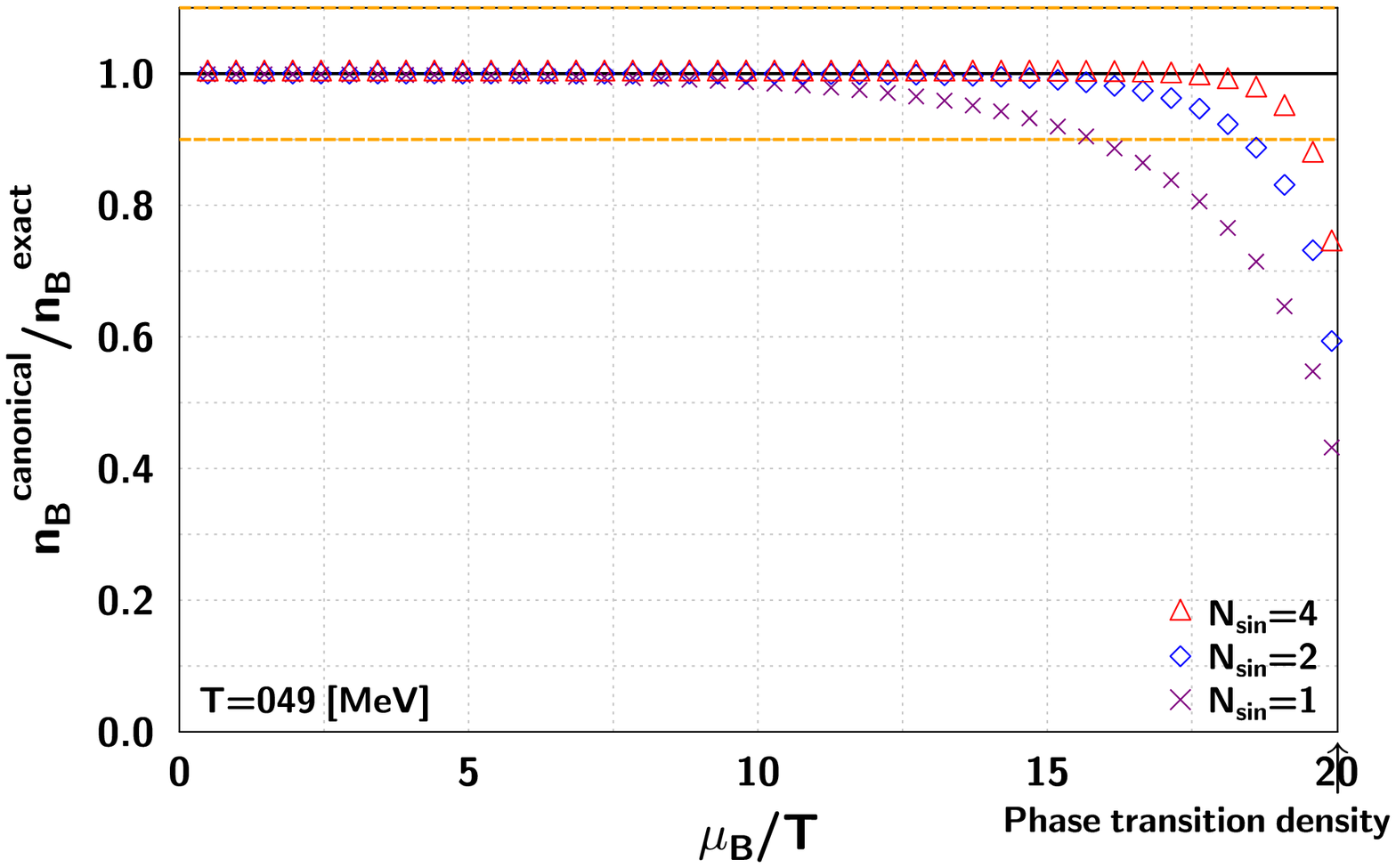}
\includegraphics[scale=0.29701]{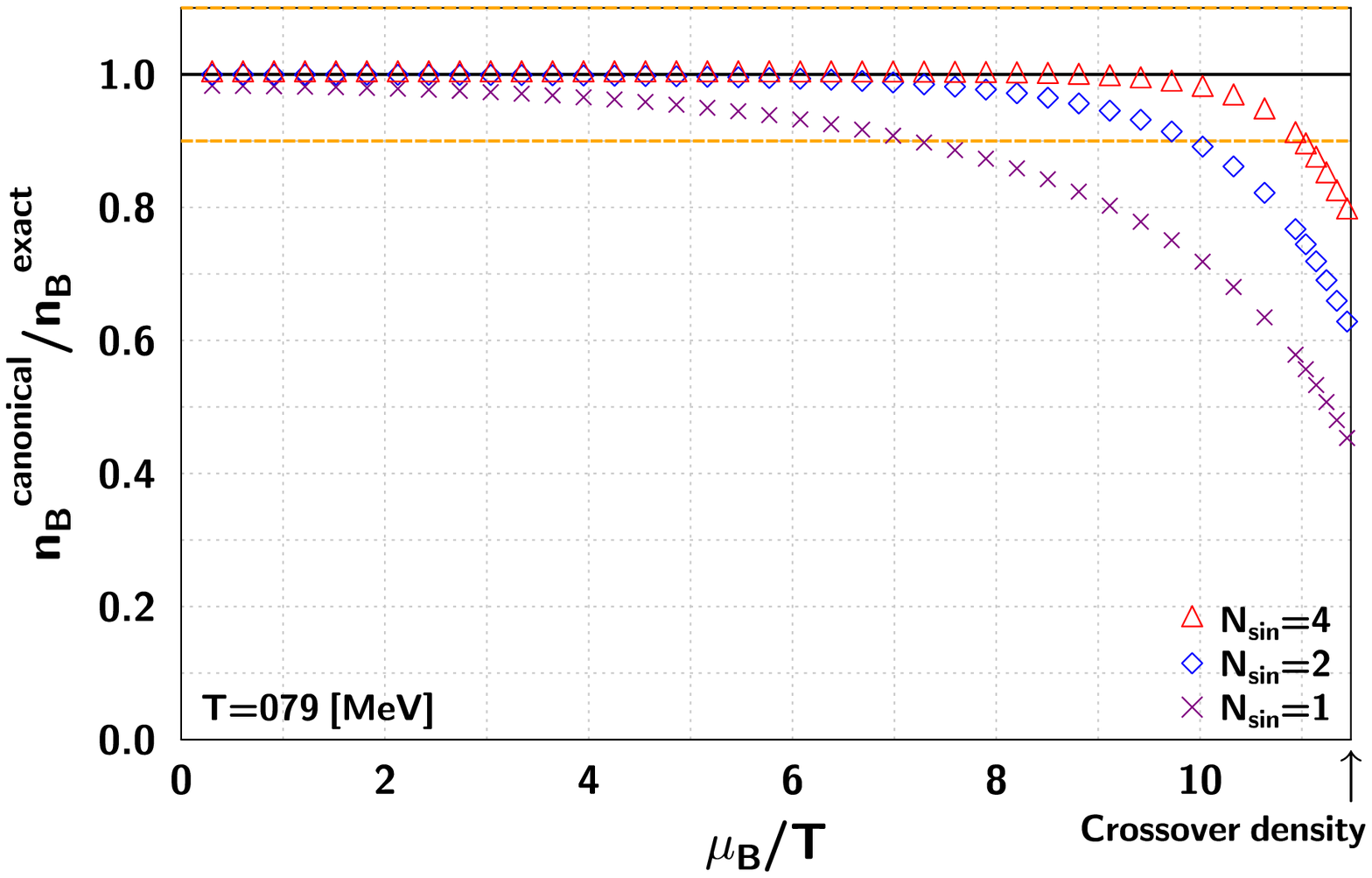}
\caption{ (color online). 
The $N_{\sin}$ dependence of $n_B^{\rm canonical}/n_B^{\rm exact}$ in the NJL model. 
$n_B^{\rm canonical}$ is the number density obtained from the canonical approach and 
$n_B^{\rm exact}$ is the exact number density calculated at the real chemical potential. 
}
\label{nsindep2NJL}
\end{center}
\end{figure*}

\begin{figure}
\begin{center}
\includegraphics[scale=0.72]{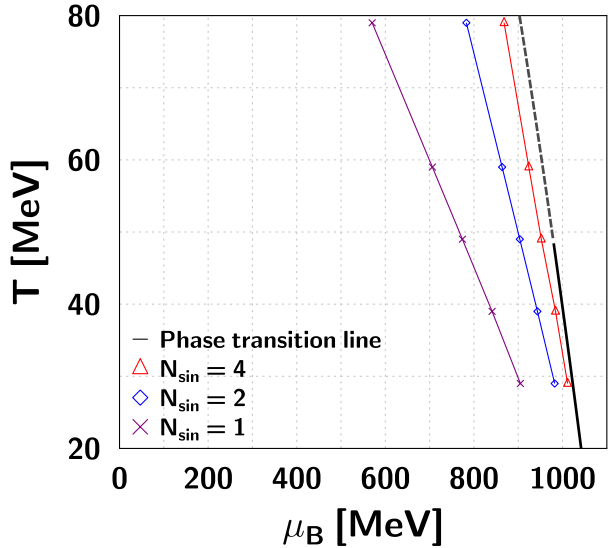}
\caption{ (color online). 
The boundaries of the effective regions of the canonical approach for $N_{\sin}$ in the NJL model. 
The black solid and dashed lines represent the first-order phase transition and crossover lines, respectively. 
The points of making a 10\% difference between the exact $n_B$ and the results from the canonical approach are plotted. 
}
\label{nsindep3NJL}
\end{center}
\end{figure}

\section{\label{summ}Summary}

We have investigated the effective region of the canonical approach in the NJL and PNJL models. 
We have calculated the 161 data of the imaginary number densities as functions of the pure imaginary chemical potential. 
By using the integration method of a Fourier series with finite $N_{\sin}$ for the imaginary number densities 
and performing Fourier transforms with the multiple-precision arithmetic, 
we have reconstructed the grand canonical partition function, 
which is written as a fugacity expansion with finite $N_{\rm max}$.  
After that, we have calculated the number densities at the real chemical potential from the grand canonical partition function. 
Because the number densities are already known in the NJL and PNJL models, 
we can clarify the region where the canonical approach works well 
by comparing the number densities obtained from the canonical approach with the exact ones. 

We have shown the $N_{\rm max}$ and $N_{\sin}$ dependences of the number densities obtained from the canonical approach 
in each model. 
In the investigation of the $N_{\rm max}$ dependence, 
we have found that 
the finite $N_{\rm max}$ effect for the number density is suppressed for 
the maximum value of the fluctuation of the net quark number density in the system, $N_{\rm max}/V$, larger than 0.56~[fm$^{-3}$]. 

For the $N_{\sin}$ dependence, 
we have found that 
the results for $N_{\sin}$ up to 4 can reconstruct the exact number density from the canonical approach 
until 96\% of the phase transition or crossover density within the 10\% difference. 
Moreover, $N_{\sin}=1$ or 2 is enough to reconstruct the exact number density within the 10\% difference 
for $T < T^\mathrm{CEP}$ and $\mu_B < 900$~[MeV]. 
The results have universality for at least the NJL and PNJL models. 
They suggest that the application of the canonical approach to the lattice QCD is useful, especially in the confinement phase. 

In this paper, we have discussed the effective region of the canonical approach for the number density in the NJL and PNJL models. 
It remains to be investigated for other physical quantities and other models.

\begin{acknowledgments}

This work was supported by the National Research Foundation of Korea (NRF) 
grant funded by the Korean government (MSIT) (2018R1A5A1025563). 
The work of SiN is also supported in part by the NRF fund (2019R1A2C1005697). 
AH is supported in part by Grants-in-Aid for Scientific Research (No.~JP17K05441 (C)) 
and for Scientific Research on Innovative Areas (No.~18H05407). 
This work was supported by 
``Joint Usage/Research Center for Interdisciplinary Large-scale Information Infrastructures" and 
``High Performance Computing Infrastructure" in Japan (Project ID: jh190051-NAH). 
The calculations were carried out on 
SX-ACE and OCTOPUS at RCNP/CMC of Osaka University.

\end{acknowledgments}


\end{document}